\documentclass{aa}
\usepackage{times}
\usepackage[sectionbib,authoryear]{natbib}
\usepackage{graphicx} % when using Latex and dvips
\begin{document}

\title {Multiband Emission from Pulsar Wind Nebulae: A Possible Injection
Spectrum}
\titlerunning{Multiband Emission from Pulsar Wind Nebulae}
\subtitle{}
\author{ J. Fang \and L. Zhang }
\offprints{L. Zhang}
\institute{Department of Physics, Yunnan University, Kunming, China\\
              \email{lizhang@ynu.edu.cn}
}

\date{Received month day, year; accepted month day, year}

\abstract{} {A recent research shows that particles with a spectrum
of a relativistic Maxwellian plus a high-energy tail can be
accelerated by relativistic collisionless shocks. We investigate the
possibility of the high-energy particles with this new spectrum
injected in pulsar wind nebulae (PWNe) from the terminate shock
based on the study of multiwavelength emission from PWNe.} {The
dynamics of a supernova remnant (SNR) and multiband nonthermal
emission from the PWN inside the remnant are investigated using a
dynamical model with electrons/positrons injected with the new
spectrum. In this model, the dynamical and radiative evolution of a
pulsar wind nebula in a non-radiative supernova remnant can be
self-consistently described.} {This model is applied to the three
composite SNRs, G0.9+0.1, MSH 15-52, G338.3-0.0, and the multiband
observed emission from the three PWNe can be well reproduced.} {Our
studies on the three remnant provide evidence for the new spectrum
of the particles, which are accelerated by the terminate shock,
injected into a PWN.}


\authorrunning{J. Fang and L. Zhang}
\maketitle
%
%________________________________________________________________

\section{Introduction}

PWNe, which are prominent sites of high-energy emission in the
Galaxy, are powered by pulsars associated with them. A pulsar inside
a PWN loses its rotational energy through a pulsar wind composed of
magnetic flux and high-energy particles
\citep[][]{GJ69,Get07,GSZ09}. The ultra-relativistic wind flows
relativistically into a non-relativistic ejecta of the ambient
supernova remnant (SNR), which results in the PWN and a termination
shock (TS), where the plasma is decelerated and heated
\citep[][]{RC84,Vet08}. High-energy particles are injected into the
nebula from the TS, and multiband nonthermal photons with energies
ranging from radio, X-ray to $\gamma$-ray bands are emitted during
the evolution of the PWN.

Usually, multi-wavelength observational results of a PWN cannot be
well reproduced by the radiation of particles injected with a single
power-law spectrum, and a broken power-law must be employed to
better explain the observations \citep[e.g.,][]{AA96,A05a,ZCF08}.
However, the physics behind the broken power-law is unclear. For a
typical PWN, the wind from the pulsar can flow relativistically with
a lorentz factor of $\sim 10^6$. The resulting TS can accelerate
particles to relativistic energy. On the other hand, based on the
long-term two dimensional particle-in-cell simulations,
\citet[][]{Sp08} found that the particle spectrum downstream of a
relativistic shock consists of two components: a relativistic
Maxwellian and a high-energy power-law tail with an index of
$-2.4\pm0.1$. Based on this finding, with the assumption that the
high-energy particles in a PWN are injected with a spectrum of a
relativistic Maxellian plus a power-law high-energy tail, we
investigate the multiwavelength emission from PWNe to test the
possibility of particles with the new spectrum injected into the
PWNe.

The dynamical evolution of the PWN is calculated basically according
to the model in \citet[][]{GSZ09}, which can self-consistently
describe the dynamical and radiative evolution of a pulsar wind
nebula in a non-radiative supernova remnant. Different from
\citet{GSZ09}, in which a single power-law injection spectrum for
the electrons/positrons is employed to discuss the radiative
properties during different phase of the PWN, we argue in this paper
that the high-energy particles are injected with the new spectrum of
a relativistic Maxellian plus a power-law high-energy tail during
the evolution in this paper, and a kinetic equation is used to
obtain the energy distribution of the particles. We apply the model
to the PWNe in the composite SNRs, G0.9+0.1, MSH 15-52 and
G338.3-0.0, which have been observed in radio, X-rays and very
high-energy (VHE) $\gamma$-rays.

G0.9+0.1 has a 2$'$ PWN inside a 8$'$ shell in the radio band
\citep[][]{HB87}. The PWN is powered by an energetic pulsar PSR
J1747-2809, which was recently discovered in G0.9+0.1 with the NRAO
Green Bank Telescope at 2GHz \citep[][]{Cet09}. A jet-like feature
of the nebula was revealed in the high-resolution observation with
{\it Chandra} \citep[][]{GPG01}. The observation in the TeV band
with HESS indicates that the PWN is a weak emitter in  VHE
$\gamma$-rays \citep[][]{A05a}.

MSH 15-52 is a complex SNR with a ragged shell in the radio
observations \citep[][]{Cea81}. An elongated PWN powered by an
energetic pulsar was found in the remnant in the X-ray observations
with ROSAT \citep[][]{Tea96}, BeppoSAX \citep{M01} and Chandra
\citep[][]{G02}. The remnant has also been well detected in VHE
$\gamma$-rays with HESS \citep[][]{A05b} and CANGAROO
\citep[][]{Nea08}, and the significant VHE $\gamma$-rays are
identified to be produced from the PWN.

A VHE $\gamma$-ray source HESS J1640-465 was discovered by HESS in
the survey of the inner Galaxy \citep[][]{A06}, and it is spatially
coincident with the composite SNR G338.3-0.0. An extended X-ray
source was found to be located at the center of the VHE $\gamma$-ray
source with XMM-Newton \citep[][]{Fea07}. Very recently,
\citet[][]{Lea09} presented the high resolution X-ray observations
on the PWN, and a point like source as a putative pulsar appears in
the X-ray observations.

In this paper, we investigate the possibility of the particles with
the new spectrum injected in PWNe based on applications with the
spectrum to the three composite SNRs. In Section \ref{sec:model},
the model is simply described, and the results from the applications
to the three SNRs G0.9+0.1, MSH 15-52 and G338.3-0.0, are presented.
The main conclusions and discussion are given in Section
\ref{sec:discussion}.

\section{Model and Results}
\label{sec:model}

A PWN is powered by the its pulsar which dissipates the rotational
energy into the nebula. The spin-down luminosity of a neutron star
with a rotation period of $P$ evolves with time as
\citep[e.g.,][]{GS06,S08}
\begin{equation}
\dot{E}(t)=\dot{E}_0(1+t/\tau_0 )^{-(n+1)/(n-1)}, \label{Lum}
\end{equation}
where $\tau_0$ is the spin-down time scale of the star, $\dot{E}_0$
is the initial spin-down power, $n$ is the braking index of the
pulsar, which is equal to 3 for magnetic dipole spin-down.

Magnetic field and high-energy particles are injected at the TS
located where the ram pressure of the unshocked wind is equal to
that of the PWN. In this paper, we assume the spin-down power is
distributed between electrons and positions ($\dot{E}_{\rm e}(t) =
\eta_{\rm e}\dot{E}(t)$), and magnetic fields ($\dot{E}_{\rm B}=
\eta_{\rm B}\dot{E}(t))$ \citep[e.g.,][]{GSZ09}. In
\citet[][]{GSZ09}, they used a single power-law injection spectrum
for the electrons/positrons to discuss the radiative properties
during different phases of the PWN evolution. However, a broken
power-law spectrum is usually needed to reproduce the non-thermal
emission of a PWN with multi-band observations
\citep[e.g.,][]{AA96,Vd06,Set08,ZCF08}. Recently, based on the
long-term two-dimensional particle-in-cell simulations,
\citet[][]{Sp08} found that the particle spectrum downstream of a
relativistic shock can be fitted as a Maxwellian plus a power-law
tail. The form is
$f(\gamma)=C_1\gamma\exp(-\gamma/\Delta\gamma_1)+C_2\gamma^{-\alpha}\min[1,\exp(-\gamma-\gamma_{\rm
cut})/\Delta\gamma_{\rm cut}]$, where $\gamma$ is the Lorentz factor
of the particles, $\Delta\gamma_1$ is close to $(\gamma_0-1)/2$,
$\gamma_0$ is the Lorentz factor of the upstream flow, $C_2=0$ for
$\gamma<\gamma_{\rm min}$, $\gamma_{\rm min}$ is about seven times
of $\Delta\gamma_1$ \citep[][]{Sp08}. In this paper, we investigate
the possibility of particles with this spectrum injected in PWNe by
multiwavelength studies.

We argued that high-energy particles injected in a PWN are
accelerated by the  TS, for which the upstream pulsar wind typically
can have a Lorentz factor $\sim10^6$. Therefore, we assume the
spectrum of the high-energy particles injected in the PWN has the
form,
\begin{equation}
Q(E,t) = \left\{\begin{array}{ll} C(t)\frac{E}{E_{\rm b}}\exp\left(-\frac{E}{E_{\rm b}}\right) & E\leq E_{\rm min}\\
C(t)\left[\frac{E}{E_{\rm b}}\exp\left(-\frac{E}{E_{\rm
b}}\right)+f\left(\frac{E}{E_{\rm b}}\right)^{-\alpha}\right] &
E_{\rm min}<E \leq E_{\rm max}
\end{array}
\right . , \label{dnde}
\end{equation}
where, $\alpha=2.4\pm0.1$, $E_{\rm b}\sim2.6\times10^5 \gamma_{\rm
ts, 6}$ MeV, $\gamma_{\rm ts, 6}$ is the Lorentz factor of the
upstream pulsar wind of the TS in units of $10^6$, $E_{\rm min} =
f_{\rm min}E_{\rm b}$ with $f_{\rm min}\sim7$, $f$ is normalized by
$E_{\rm min}/E_{\rm b}\exp\left(-E_{\rm min}/E_{\rm
b}\right)=f\left(E_{\rm min}/E_{\rm b}\right)^{-\alpha}$. This
spectrum is similar to that given in \citet{Sp08} except for a
discontinuity at $E_{\rm min}$, which has no influence on the final
multiband photon spectrum, and the cut-off form at the highest
energy. We leave $E_{\rm max}$ as a parameter constrained by the
observational results for a PWN.  $C(t)$ can be obtained with
\begin{equation}
C(t) = \frac{\dot{E}_{\rm e}(t)}{2E_{\rm b}^2 + f\frac{E_{\rm
b}^2}{2-\alpha} \left[ \left( \frac{E_{\rm max}}{E_{\rm b}}
\right)^{2-\alpha} - \left( \frac{E_{\rm min}}{E_{\rm b}}
\right)^{2-\alpha}\right]}. \label{C}
\end{equation}
Assuming the particles are homogeneous distributed in space in the
PWN, and the distribution in the emission region is given by
\begin{equation}
\frac{\partial N(E,t)}{\partial t}= \frac{\partial }{\partial E}
\left [ \dot{E} N(E,t) \right ] + Q(E,t) , \label{Distr}
\end{equation}
where $\dot{E}$ is the energy-loss rate of the particles with energy
$E$. Energy-loss mechanisms include synchrotron radiation, inverse
Compton scattering and adiabatic losses.

The dynamics of the PWN in the supernova shell is calculated
basically following the model presented in \citet[][]{GSZ09}. The
model assumes that the progenitor supernova ejects material with
mass $M_{\rm ej}$ and energy $E_{\rm sn}$ into the ambient matter
with a constant density $\rho_{\rm ISM}$. Assuming the PWN has no
influence on the dynamics of the forward shock and the reverse
shock, the velocity and the radius of the forward shock and the
reverse shock of the surrounding SNR are calculated with the
equations in \citet[][]{TM99}. The pulsar dissipates energy into the
PWN, which sweeps up the supernova ejecta into a thin shell
surrounding the nebula, and new particles are injected into the PWN
at each time step.

We now apply the model the investigate the three composite SNRs,
G0.9+0.1, MSH 15-52 and G338.3-0.0. Firstly, if the distance, the
age and the radii of the SNR shell and the PWN are known, the
parameters such as the supernova energy ($E_{\rm sn}$), the ejecta
mass ($M_{\rm ej}$), the ambient density ($\rho_{\rm ISM}$), and the
initial spin-down energy ($\dot{E_0}$) can be constrained by making
the results consistent with the known values. In our calculation, a
typical spin-down time scale ($\tau_{\rm sd}$) is set to 500 yr for
the three remnants, and $\dot{E_0}$ is also constrained by the
current spin-down power of the pulsar if it is available. Finally,
in order to obtain a consistent nonthermal emmission with the
multiband observational fluxes for the SNR, the other parameters can
be constrained.  Note that uncertainties for some parameters still
exist due to both their insensitivities on the final results and the
uncertainties of the key properties of the system, such as the
distance, the age and the properties of the pulsar.

\subsection{G0.9+0.1}
\label{Sec:appg09}

\begin{table*}
\begin{center}
\caption{Input Parameters for the three PWNe \label{para}}
\begin{tabular}{lcccc}
\hline \hline
{\sc Parameter} & {\sc G0.9+0.1} & {\sc MSH 15-52} & {\sc G338.3-0.0 (case 1)} & {\sc G338.3-0.0 (case 2)} \\
\hline
Distance (kpc)              & 8.5             & 5.2         &  10.0   & 10.0    \\
$T_{\rm age}$ (yr)          & 1900            & 1900        &  4500   & 8200    \\
$\tau_{\rm sd}$ (yr)        & 500             & 500         &   500   & 500     \\
$M_{\rm ej}$ ($M_{\odot}$)  & 8.0             & 8.0         &   8.0   &  8.0    \\
$E_{\rm sn}$ ($10^{51}$erg) & 1.0             & 3.0         &   0.5   & 0.8     \\
$n_{\rm ism}$ (cm$^{-3}$)   & 0.01            & 0.01        &   0.1   & 1.0     \\
$\dot{E_0}$ ($10^{39}$ erg s$^{-1}$)  & 2.0   &  1.5        &   10    & 0.5     \\
$n$                         & 3.0             & 2.84         &   3.0  & 3.0     \\
$\eta_{\rm B}$ ($10^{-3}$)  & 1.0             &  20.0        &   1.0  & 0.3     \\
$\alpha$                    & 2.5             &  2.4        &  2.5    & 2.5     \\
$E_{\rm max}$ (TeV)         & 900             &  1000       &  500    & 500     \\
$E_{\rm b}$  ($10^5$MeV)    & 0.2             &  0.9        &  0.8    & 0.8     \\
\hline
\end{tabular}
\end{center}
\end{table*}

\begin{figure}
\begin{center}
\includegraphics[scale=1.2]{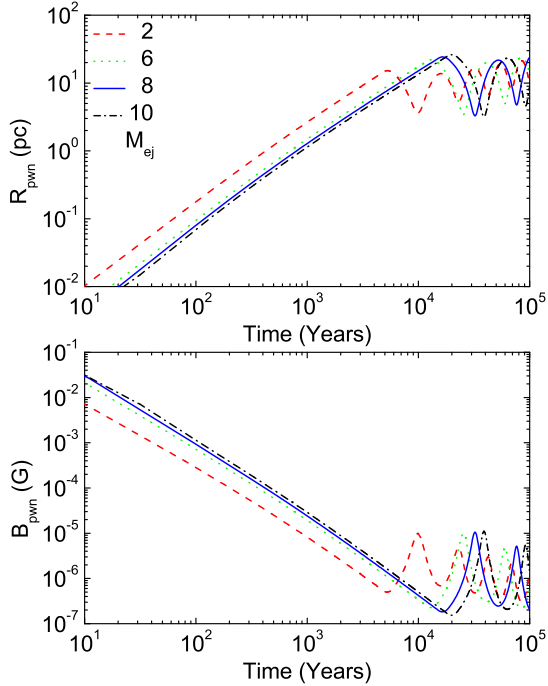}
\caption{\label{Figs1} Upper panel: radius of PWN ($R_{\rm pwn}$)
for $M_{\rm ej}=2$ (dashed line), $M_{\rm ej}=6$ (dotted line),
$M_{\rm ej}=8$ (solid line) and $M_{\rm ej}=10$ (dash-dotted line)
$M_{\odot}$, respectively, with the other parameters listed in Table
\ref{para} for G0.9+0.1. Lower panel: magnetic field strength in the
PWN ($B_{\rm pwn}$) with different $M_{\rm ej}$, and the others are
same as the upper panel. }
\end{center}
\end{figure}

The composite SNR G0.9+0.1 consists of a bright 2$'$ PWN hosted by a
8$'$ shell in the radio band \citep[][]{HB87,Let00}. Recent high
resolution radio study of the PWN indicates that the radio emission
of the nebula has a spectral index of $-0.18\pm0.04$, and the fluxes
are 1.35, 1.45 and 1.75 Jy for the wavelength 3.6, 6 and 20 cm,
respectively \citep[][]{DGD08}. Furthermore, an energetic pulsar PSR
J1747-2809 with a period of 52 ms has recently discovered in
G0.9+0.1 with the NRAO Green Bank Telescope at 2GHz
\citep[][]{Cet09}.

The PWN in G0.9+0.1 was detected in X-rays with BeppoSAX, and the
result shows that the X-ray emission from the source has a power-law
spectrum with a photon index of $2.0\pm0.3$ \citep[][]{MSI98,Set00}.
High-resolution observation with Chandra indicated that the nebula
is axisymmetric with a jet-like feature, which is evidence of a
torus of emission in the pulsar's equatorial plane and a jet
directed along the pulsar spin axis \citep[][]{GPG01}.
\citet[][]{PDW03} studied G0.9+0.1 using observations by {\it
XMM-Newton}. The X-ray spectrum softens with distance from the core,
and the spectrum in the energy range 2 -- 10 keV has a power law
form with a photon index of $\sim1.9$.

Very High-energy (VHE) emission from G0.9+0.1 has been detected with
HESS \citep[][]{A05a}. The photon flux above 0.2 TeV is
$5.7\times10^{-12}$ cm$^{-2}$ s$^{-1}$, and the spectrum can be
fitted with a power-law with a photon index $2.4\pm0.31$. The source
is a weak TeV emitter, and the VHE $\gamma$-rays appear to originate
from the core rather than the shell \citep[][]{A05a}.

The dynamical and radiative properties of the composite SNR G0.9+0.1
are investigated with the parameters in Table \ref{para} for this
source. Although the pulsar PSR J1747-2809 has a characteristic age
of 5.3 kyr, \citet[][]{Cet09} argued that G0.9+0.1 has a small age
of no more than $2-3$ kyr either from PWN evolution models of
\citet[][]{BCF01} for the observed ratio $R_{\rm pwn}/R_{\rm snr} =
0.25$ or from the PWN energetics \citep[][]{DGD08}. The distance of
the pulsar is likely in the range of 8 kpc to 16 kpc due to the
uncertainty of the electron density model toward the distant inner
Galactic regions \citep[][]{DGD08}. We assume the distance is 8.5
kpc in the calculation, then the radii of the PWN and the shell are
2.55 pc and 10.2 pc, respectively. Moreover, with $E_{\rm
sn}=10^{51}$ erg and $M_{\rm ej}=8M_{\odot}$, an age of 1900 yr and
a relatively low density $n_{\rm ism}=0.01$ cm$^{-3}$ are needed to
well reproduce structure of the system, i.e., the radius of the
shell and the ratio $R_{\rm pwn}/R_{\rm snr}$. A pulsar's velocity
of 120 km/s similar as the Crab pulsar \citep[][]{Ket08} is used to
illustrate the evolution of the PWN, which has no influence on the
final results for G0.9+0.1 since it is a young remnant, and now the
pulsar is safely in the nebula. With these parameters, the resulting
radii of the PWN and the SNR shell are 2.6 and 10.2 pc,
respectively, consistent with the observational results (upper panel
Fig.\ref{Figs1}). The influence of the ejecta mass $M_{\rm ej}$ on
the radius of the nebula is shown in the upper panel of
Fig.\ref{Figs1}. With a smaller $M_{\rm ej}$, both the nebula and
the SNR shell expand more quickly, and the nebula collides with the
reverse shock at later time. As a result, the magnetic field in the
PWN is weaker for a smaller $M_{\rm ej}$ due to the bigger volume of
the nebula (the lower panel in Fig.\ref{Figs1}).

\begin{figure*}
\begin{center}
\includegraphics[scale=1.2]{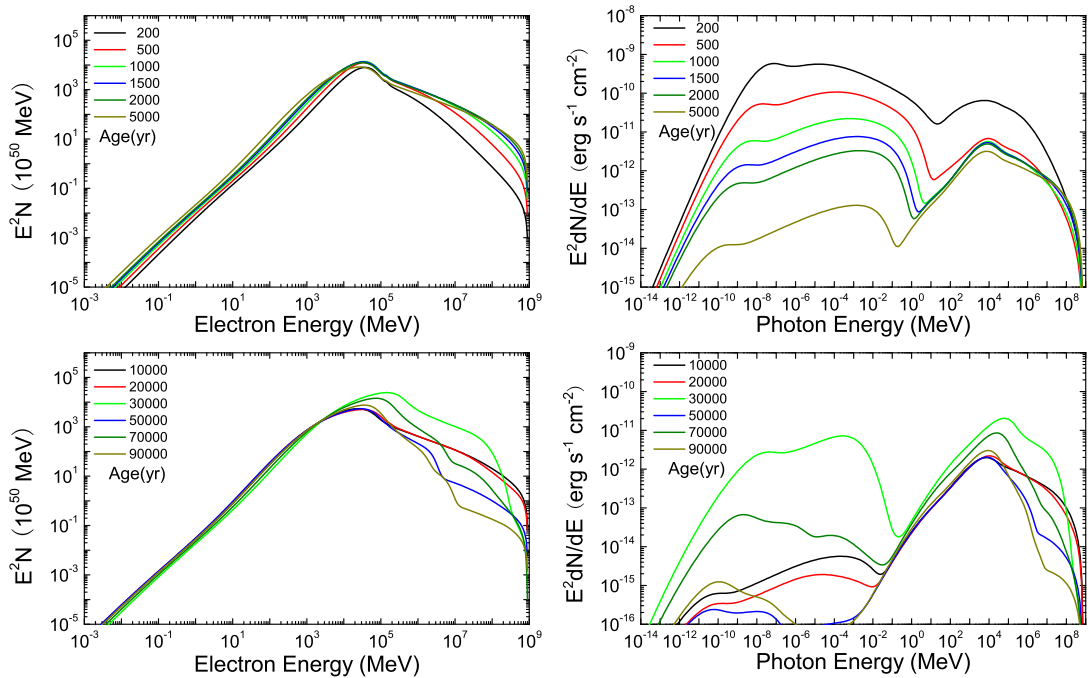}
\caption{\label{Figspe} Left panels: particle spectra of electrons
and positrons at different times during the evolution of the PWN.
Right panels: the resulting spectral energy distribution at
different ages with the other parameters listed in Table.\ref{para}.
}
\end{center}
\end{figure*}

The dynamical properties of the SNR during the evolution is similar
as those for the Crab remnant in \citet[][]{GSZ09} although the new
spectrum of the particles is used in this paper. Initially, the
pressure of the PWN is much bigger that of the surrounding supernova
ejecta, so the PWN expands adiabatically into the cold supernova
ejecta. The ejecta surrounding the PWN is swept up to a thin shell,
which is decelerated by ram pressure since its velocity is higher
than the local sound speed \citep[][]{Get07}. The mass of the PWN
$M_{\rm pwn}$ increases continuously since the shell expands faster
than the ambient ejecta. This expansion phase ends when the PWN
collides with the reverse shock of the SNR. After the collision, the
pressure of the nebula $P_{\rm pwn}$ is much less than that of the
supernova ejecta around the nebula $P_{\rm snr}(R_{\rm pwn})$. The
velocity of the PWN decreases greatly, and finally the PWN is
compressed when the shell moves inwardly. During this process of
compression, the magnetic field strength in the nebula increases
significantly (lower panel in Fig.\ref{Figs1}), and, as a result,
the synchrotron luminosity has a rapid rise (lower right panel in
Fig.\ref{Figspe}). Furthermore, the radius of the PWN decreases
significantly, and the PWN will expand again when the pressure in it
eventually becomes bigger than that of the surrounding ejecta. The
nebula enjoys a series of contractions and re-expansions until the
SNR enters the radiative phase of its evolution. The pulsar with a
velocity moving in the space will leave the PWN if the velocity is
high enough in the compression process, and it can re-enter the
nebula when the nebula expands again.

Particles with a spectrum of a Maxwellian plus a power-law tail
($\alpha=2.5,\, f=1.0$) are injected in the PWN when the pulsar is
inside it. The particles lose energy through synchrotron radiation,
inverse Compton scattering and adiabatic loss. Non-thermal emission
from the PWN from the radio to X-ray band is produced via
synchrotron radiation, whereas $\gamma$-rays are produced through
inverse Compton scattering on the seed photons, i.e., cosmic
microwave background (CMB), infrared (IR) and optical (opt) photons.
For the soft seed photons, the energy density of the infrared
radiation is 0.23 eV cm$^{-3}$ as used for the GALPROP code
\citep{SMR00,Pea06} and a density of 5.7 eV cm$^{-3}$ for the
optical component, which is 50\% smaller than the value used in
GALPROP, are used in \citet[][]{A05a} to discuss the origin of the
VHE $\gamma$-rays. We find densities of 0.5 eV cm$^{-3}$ for the
infrared component ($kT_{\rm IR}=3\times10^{-3}$ eV) and 20 eV
cm$^{-3}$ for the optical soft photons ($kT_{\rm opt}=0.25$ eV) can
reproduce the observed spectrum in the VHE $\gamma$-ray band well,
so these densities are used in this paper. The resulting energy
distribution of the particles in the PWN and the multi-wavelength
non-thermal emission during the evolution are presented in
Fig.\ref{Figspe}. The particles with energies smaller than $E_{\rm
b}$ are mainly from the high-energy particles with energies $>E_{\rm
b}$ experienced energy losses, and the energy distributions in the
range $<E_{\rm b}$ at different age in the evolution all show a
power-law form with an index of $\sim 1$ ($dN/dE\propto E^{-1}$). On
the other hand, the energy distribution in the higher-energy band
($>E_{\rm b}$) becomes steady between 1000 yr and 20000 yr, and the
distribution can also be represented as a power-law but with an
index $\sim 2.5$.

The expansion and compression of the PWN cause the magnetic field
strength in the nebula $B_{\rm pwn}$ to decrease and increase,
respectively. The synchrotron emissivity diminishes before 10000 yr
as the PWN expands into the cold supernova ejecta, when the magnetic
field strength in the nebula decreases gradually. However,
synchrotron radiation can be important again when the PWN is
compressed, because the magnetic field strength rises greatly during
the compression. In this case, the energy of the particles in the
nebula can have a rise due to the adiabatic compression, then the
emissivity of inverse Compton scattering can also have a rise (see
the emission in Fig.\ref{Figspe} for 30000 yr). During the whole
evolution process, the PWN is an important emitter of the GeV
$\gamma$-rays, although the emission from the radio to the X-ray
band is insignificant sometimes.

\begin{figure}
\begin{center}
\includegraphics[scale=0.7]{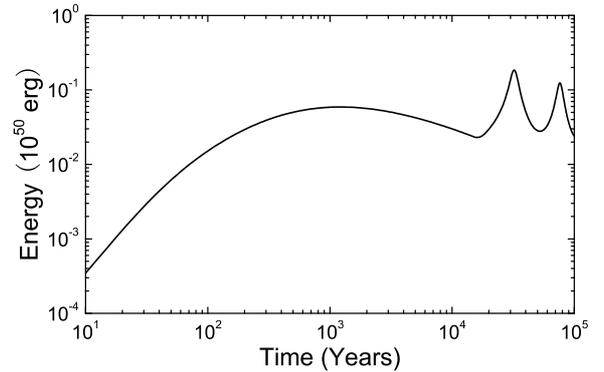}
\caption{\label{Epwn} The energy contained in the PWN at different
ages of the system with the parameters same as Fig.\ref{Figspe}.}
\end{center}
\end{figure}

\begin{figure}
\begin{center}
\includegraphics[scale=0.7]{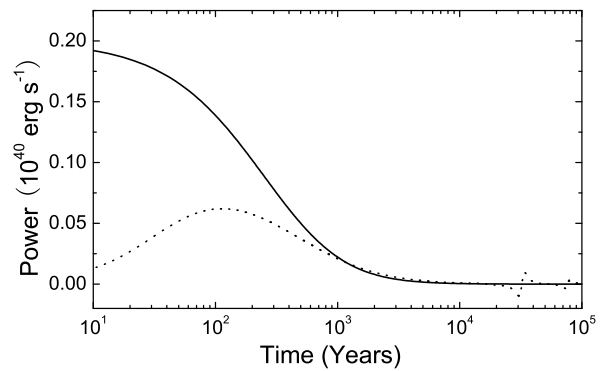}
\caption{\label{Epower} The spin-down power of the pulsar (solid
line) and the adiabatic loss rate of the PWN (dotted line) during
the evolution of the system. The others are the same as
Fig.\ref{Epwn}.}
\end{center}
\end{figure}

Fig.\ref{Epwn} shows the evolution of the energy contained in the
PWN ($E_{\rm pwn}$) during the evolution of the system, and the
spin-down power of the pulsar and the adiabatic loss rate of the
PWN, which can be obtained with $L_{\rm ad}=E_{\rm pwn}v_{\rm
pwn}/R_{\rm pwn}$, are indicated in Fig.\ref{Epower}. Here $v_{\rm
pwn}$ and $R_{\rm pwn}$ are the velocity and radius of the PWN,
respectively. At first, the energy of PWN increases continuously
since the pulsar injects the spin-down power into the nebula, and
this stage ends after $\sim1000$ yr (Fig.\ref{Epwn}) when the
adiabatic loss rate is considerable compared with the spin-down
power of the pulsar (Fig.\ref{Epower}). After this phase, the energy
of the nebula decreases or increases due to the competition between
the spin-down power of the pulsar adiabatic loss.

\begin{figure}
\begin{center}
\includegraphics[scale=1.2]{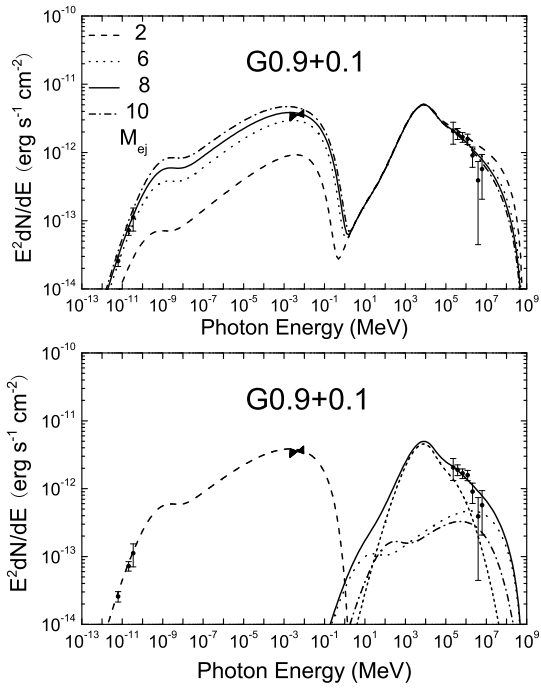}
\caption{\label{Figspe2} Upper panel: the multiband spectral
distribution of the nonthermal emission for $M_{\rm ej}=2$ (dashed
line), $M_{\rm ej}=6$ (dotted line), $M_{\rm ej}=8$ (solid line) and
$M_{\rm ej}=10$ (dash-dotted line) $M_{\odot}$ with the other
parameters listed in Table \ref{para} for G0.9+0.1. Lower panel: the
resulting photon emission of synchrotron (dashed line), inverse
Compton scattering on the CMB (dotted line), IR(dash-dotted line),
starlight (short-dashed line), and synchrotron photons (close-dotted
line) for $M_{\rm ej}=8M_{\odot}$ and the others same as the upper
panel. The solid line represents the whole inverse Compton
scattering of the electrons/positions on all the soft photons. The
radio \citep[][]{DGD08}, X-ray \citep[][]{PDW03} and VHE
$\gamma$-ray \citep[][]{A05a} observations for the PWN are also
shown in the figure for comparison. }
\end{center}
\end{figure}

\begin{figure}
\begin{center}
\includegraphics[scale=0.7]{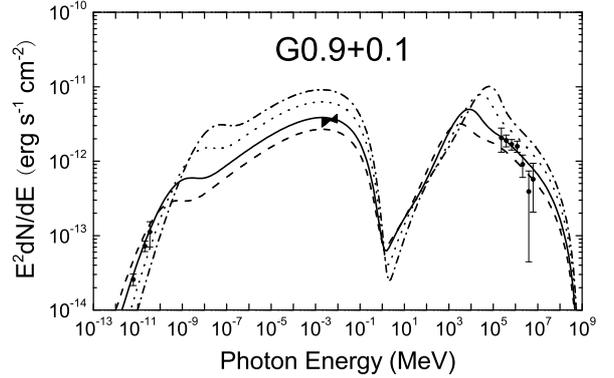}
\caption{\label{Figspedeb} The resulting spectral energy
distributions for $E_{\rm b}=0.1\times10^5$ (dashed line),
$0.2\times10^5$ (solid line), $0.5\times10^5$ (dotted line) and
$1.0\times10^5$ (dash-dotted line) MeV, respectively. The others are
the same as the lower panel in Fig.\ref{Figspe2}. }
\end{center}
\end{figure}

With the parameters in Table.\ref{para}, the dynamical structure and
the multi-band radiative properties of G0.9+0.1 are well reproduced
at 1900 yr. At this age, the magnetic field strength in the PWN is
8.1 $\mu$G, and the current spin-down power of the pulsar is
$8.67\times10^{37}$ erg s$^{-1}$, which is still consistent with the
measured one $4.3\times10^{37}$ erg s$^{-1}$ \citep[][]{Cet09} due
to the uncertainty of the moment of inertia; moreover, the
observational radii of the nebula and the SNR shell are reproduced
in this scenario. The resulting multi-band emission of the PWN from
the model is shown in the lower panel in Fig.\ref{Figspe2}. The
radio and the X-ray observations can be well explained as
synchrotron radiation of the particles injected in the PWN; VHE
$\gamma$-rays from the nebula are mainly produced by inverse Compton
scattering on the optical light and the cosmic microwave background.

With a bigger ejecta mass ($M_{\rm ej}$), the radius of the PWN is
smaller due to the deceleration of the ejecta matter before the
collision between the reverse shock and the nebula. As a result, the
resulting  synchrotron radiation is stronger for a bigger $M_{\rm
ej}$ (the upper panel in Fig.\ref{Figspe2}). Moreover, the influence
of the break energy $E_{\rm b}$ on the resulting multiband
nonthermal emission is indicated in Fig.\ref{Figspedeb} for $E_{\rm
b}=0.1\times10^5$ (dashed line), $0.2\times10^5$ (solid line),
$0.5\times10^5$ (dotted line) and $1.0\times10^5$ (dash-dotted line)
MeV, respectively. With a bigger $E_{\rm b}$, the resulting fluxes
in X-rays and $\gamma$-rays are higher. The multiband observational
results in the radio, X-rays and VHE $\gamma$-rays can be well
reproduced with $E_{\rm b}=0.2\times10^5$. In such a case, the
Lorentz factor of the pulsar wind upstream of the TS can be
estimated as $\sim 0.8\times10^5$.

\subsection{MSH 15-52}

\begin{figure}
%\begin{center}
\includegraphics[scale=0.7]{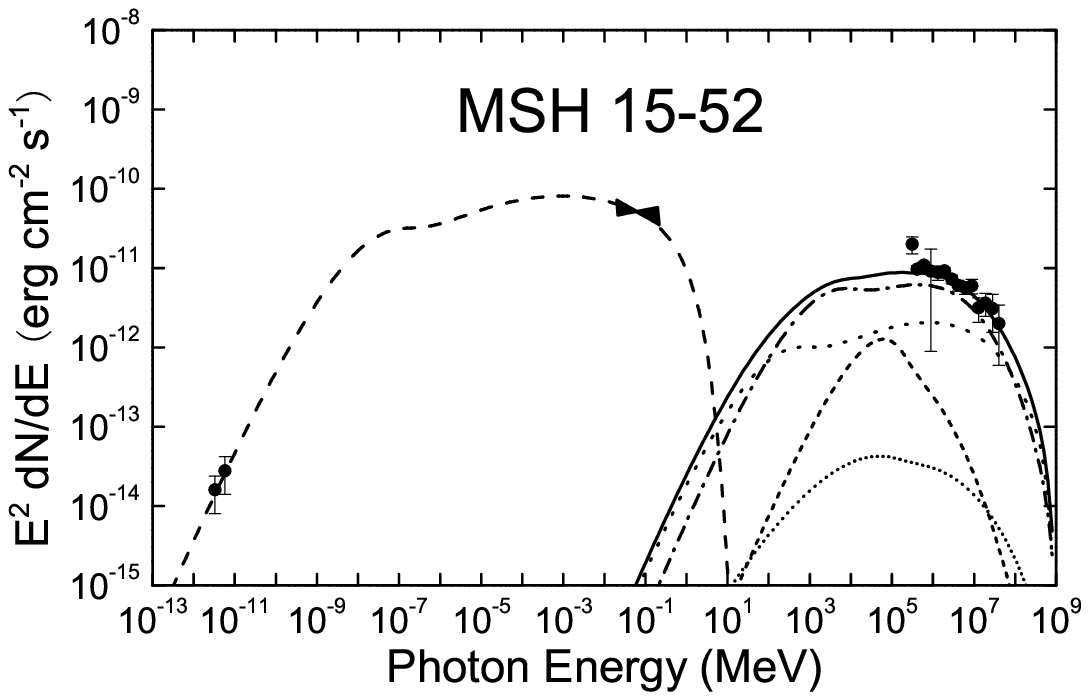}
\caption{\label{Figmsh} Comparison of the resulting photon emission
of synchrotron (dashed line), inverse Compton scattering on the CMB
(dotted line), IR(dash-dotted line), starlight (short-dashed line),
and synchrotron photons (close-dotted line) with the radio
\citep[][]{G02}, X-ray \citep[][]{M01} and VHE $\gamma$-ray
\citep[][]{A05b} observations for the PWN in MSH 15-52. The
parameters are listed in Table\ref{para} for this remnant. }
%\end{center}
\end{figure}

MSH 15-52 (G320.4-1.2) is a complex object located at a distance of
$5.2\pm1.4$ kpc based on the HI absorption measurement, consistent
with the value of $5.9\pm0.6$ kpc from the pulsar dispersion measure
\citep[][]{TC93}. It is a rough circular SNR with a diameter of
$\sim30'$ from the radio observations \citep[][]{Cea81}, and a PWN
powered by an energetic pulsar PSR B1509-58 \citep[e.g.,][]{G02} was
discovered in the remnant. \citet[][]{Lea05} presented an updated
timing solution for the young energetic pulsar PSR B1509-58 based on
the 21.3 yr of radio data and the 7.6 yr of X-ray timing data. The
results show that the pulsar has a period of about 150 ms with a
braking index of $n=2.839\pm0.003$, and a characteristic age of
$\sim 1700$ yr. In the radio band, the source appears as a
shell-like SNR with bright spots in the NW and SE region
\citep[][]{M01}, and the NW region coincides with the IR and optical
nebula RSW 89 \citep[][]{S82}. The PWN around the pulsar is
comparatively faint in radio bands, and \citet[][]{G02} estimated a
flux density of $2\pm1$ Jy at both 0.8 and 1.4 GHz using the data of
\citet[][]{WG96} and \citet[][]{G99} for the PWN. In the X-ray band,
diffuse emission is detected from the shell-like remnant with a
bright spot coincident with the NW zone. The PWN has an elongated
structure roughly centered on the pulsar with two arms extending
several arcminutes along the NW and SE directions in the X-ray
observations with ROSAT \citep[][]{Tea96} and Chandra
\citep[][]{G02}. The hard X-ray spectrum for the PWN was measured by
BeppoSAX and the power-law fit of the unpulsed high energy flux gave
a photon index $2.1\pm0.2$ and a flux of $1.2\times10^{-10}$ erg
cm$^{-2}$ s$^{-1}$ in the 20--200 keV energy range \citep[][]{M01}.
At VHE $\gamma$-ray energies, the SNR MSH 15-52 has been observed
with the HESS \citep[][]{A05b} and CANGAROO \citep[][]{Nea08}. The
TeV $\gamma$-rays are from an elliptically shaped region, and the
jet extends more prominently to the south/southeast. This morphology
coincides with the diffuse PWN as observed at the X-ray bands
\citep[][]{A05b}. The overall energy spectrum can be fitted by a
power-law with photon index
$\Gamma=2.27\pm0.03_{\rm{stat}}\pm0.02_{\rm{syst}}$ in the energy
range 280 GeV to 40 TeV in the HESS result, and a compatible result
has been obtained with the CANGAROO telescope.

The age of the SNR MSH 15-52 has been estimated as 6--20 kyr with
the standard parameters for the interstellar medium and for the
supernova explosion \citep[][]{Sea83}, which is significantly larger
than the characteristic age (1700 yr) of pulsar. However, it is very
likely that the system is young with an age of $\sim1700$ yr since
the SNR has expanded rapidly into a low-density cavity
\citep[][]{Sea83}, which is supported by the observation of HI
emission in this region \citep[][]{Dea02}.

The resulting multiband nonthermal emission from MSH 15-52 are shown
in Fig.\ref{Figmsh}, and the parameters are listed in
Table\ref{para}. With an initial spin-down power of
$1.5\times10^{39}$ erg s$^{-1}$ and $\tau_0=500$ yr, the spin-down
power of the pulsar is $5.7\times10^{37}$ erg s$^{-1}$ at an age of
1900 yr, consistent with the measured one, $5.4\times 10^{37}$ erg
s$^{-1}$ \citep[][]{Lea05}. An energy density of 1.5 eV cm$^{-3}$
for the optical soft photons is used in the calculation, which is
consistent with the value from the GALPROP code
\citep[][]{SMR00,Pea06}. For the infrared component, we find a
density of 1.5 eV cm$^{-3}$ is needed to well reproduce the
observation in VHE $\gamma$-rays, so this value is employed in the
calculation. With the parameters in Table\ref{para} for MSH 15-52,
the radii of the PWN and the SNR shell are 3.6 pc and 14.8 pc,
respectively, and the resulting magnetic field strength inside the
PWN is 19.3 $\mu$G now. The observed emission from the radio
\citep[][]{G02} to the X-ray bands detected with BeppoSAX
\citep[][]{M01} can be well explained as the synchrotron radiation
of the high-energy electrons/positrons injected in the nebula;
moreover, the resulting emission from 100 MeV to 10 TeV is mainly
produced via inverse Compton scattering off the soft infrared
photons, and the HESS flux points can be well reproduced.

\subsection{G338.3-0.0}

\begin{figure}
%\begin{center}
\includegraphics[scale=0.7]{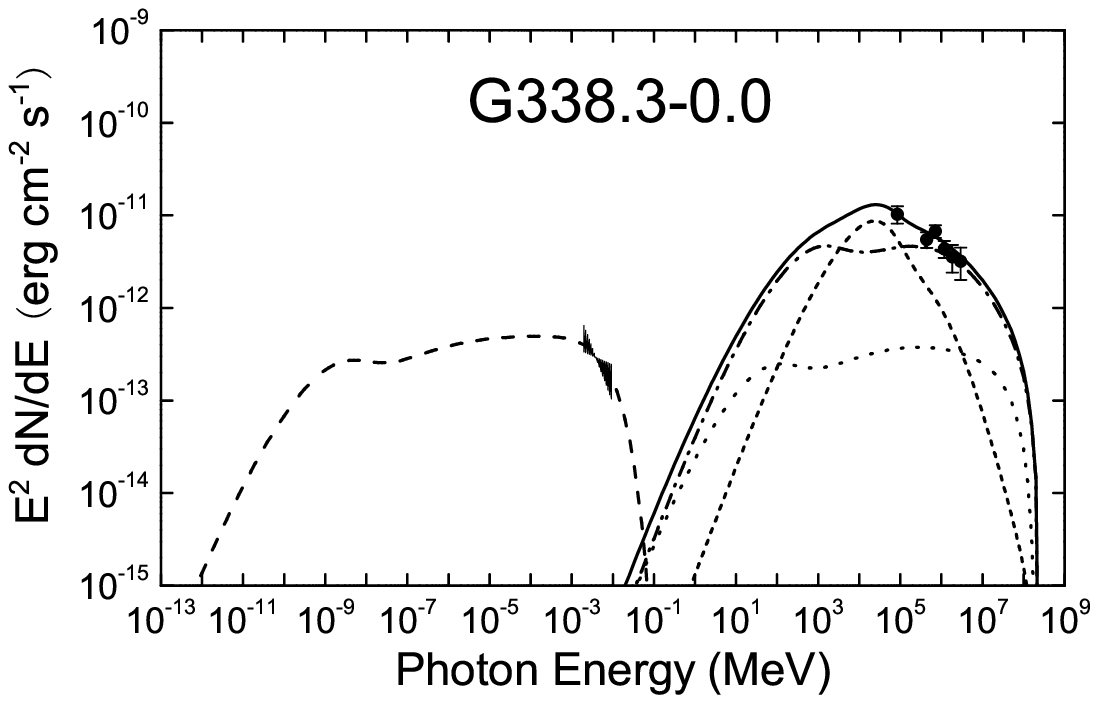}
\caption{\label{Figj1640a} Comparison of the resulting photon
emission of synchrotron (dashed line), inverse Compton scattering on
the CMB (dotted line), IR(dash-dotted line), starlight (short-dashed
line), and synchrotron photons (close-dotted line) with the X-ray
\citep[][]{Lea09} and VHE $\gamma$-ray \citep[][]{A06} observations
for the PWN in G338.3-0.0. The parameters are listed in
Table\ref{para} for G338.3-0.0 (CASE 1).}
%\end{center}
\end{figure}

HESS J1640-465 was discovered in the survey of the inner Galaxy with
HESS as a center filled VHE $\gamma$-ray source with a differential
energy spectrum fitted as a power-law with an index of $2.42\pm0.14$
above 0.2 TeV \citep[][]{A06}. This source is spatially coincident
with the SNR G338.3-0.0, which has a broken shell with a diameter of
$8'$ indicated in the the 843 MHz radio survey using the Molonglo
Observatory Synthesis Telescope \citep[][]{WG96}. A highly absorbed
source inside the shell was indicated in the X-ray observations with
ASCA \citep[][]{Sea01} and Swift \citep[][]{Lea06}. Extended X-ray
emission centered about the VHE $\gamma$-rays was detected with a
dedicated XMM-Newton observation in 2006 \citep[][]{Fea07}.
Recently, \citet[][]{Lea09} presented the high resolution X-ray
observations with Chandra on the extended source. The observed
morphology shows a PWN and a point-source presented as a potential
pulsar. The spectrum of the putative pulsar and the PWN can be
fitted with a power-law with an index of 1.1 and 2.5, respectively
\citep[][]{Lea09}. They argued that the pulsar's spin power is
$\sim4\times10^{36}$ erg s$^{-1}$ based on the X-ray luminosities of
the putative pulsar and nebula between 2 to 10 keV. The distance of
the source is less certain, and it can be from 8 kpc to 13 kpc based
on the H I absorption features observed along the line of sight
\citep[][]{Lea09}.

Assuming the SNR has a distance of 10 kpc, the radius of the radio
shell is $\sim12$ pc ($4'$) \citep[][]{WG96}, and that of the PWN is
$\sim3.5$ pc ($1.2'$) in the X-rays according to the high-resolution
observations with Chandra \citep[][]{Lea09}. Firstly, G338.3-0.0 is
assumed to be a young SNR expanding into a tenuous medium with a
density of $0.1$ cm$^{-3}$. The age of the remnant should not be too
young since the PWN must contain enough energetic particles to
produce the observational VHE $\gamma$-rays through inverse Compton
scattering. As a result, the supernova explosion energy $E_{\rm sn}$
and the initial spin-down power of the pulsar are chosen to be
10$^{51}$ erg and 10$^{40}$ erg s$^{-1}$, respectively; moreover,
E$_{\rm max}$ is constrained to 500 TeV to make the resulting
emission consist with the observations in X-rays, and relatively
high densities are needed to reproduce the VHE $\gamma$-ray fluxes
detected with H.E.S.S., i.e., 5.0 eV cm$^{-3}$ and 25.0 eV cm$^{-3}$
for the infrared and the optical soft photons, respectively. At an
age of 4500 yr, with the other parameters listed in Table\ref{para}
for the SNR G338.3-0.0 (CASE 1), the radii of the shell, the reverse
shock and the PWN are 6.9, 8.8 and 11.1 pc, respectively, and the
magnetic field strength in the nebula is 3.7 $\mu$G. In such a case,
the pulsar is energetic with a spin-down power of $1.0\times10^{38}$
erg s$^{-1}$ now.  Furthermore, the observed spectra in the X-rays
with Chandra \citep[][]{Lea09} and in the VHE $\gamma$-rays with
H.E.S.S. \citep[][]{A06} can be reproduced (Fig.\ref{Figj1640a}).
Inverse Compton scattering on the optical light dominates the
resulting emission at 0.1 TeV, whereas the emission above 1 TeV is
mainly produced via inverse Compton scattering off the IR photons.

\begin{figure}
%\begin{center}
\includegraphics[scale=0.7]{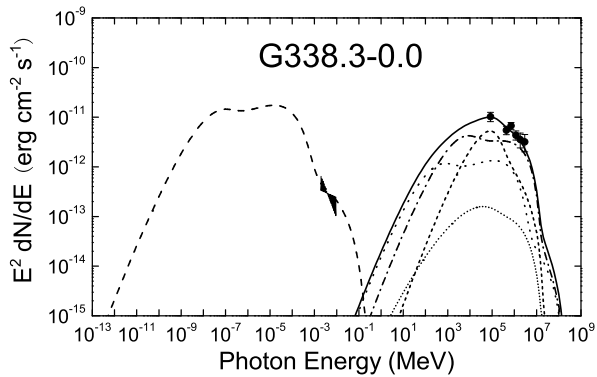}
\caption{\label{Figj1640b} The resulting spectral energy
distribution with the parameters listed in Table\ref{para} for the
CASE 2 of G338.3-0.0 (CASE 2). Others are the same as
Fig.\ref{Figj1640a}.}
%\end{center}
\end{figure}

On the other hand, we investigate the dynamical and radiative
properties of G338.3-0.0 expanding into a medium with a density of
$1$ cm$^{-3}$. With $E_{\rm ej}=0.8\times10^{51}$ erg, the radius of
the shell of the SNR in radio means that the age of the remnant is
$\sim 8000$ yr. With an initial spin-down power of
$0.3\times10^{39}$ erg s$^{-1}$ and other appropriate parameters
(see Table\ref{para} for G338.3-0.0 CASE 2), the current spin-down
power of the pulsar is $1.65\times10^{37}$ erg s$^{-1}$, and the
resulting radius of the PWN is just 2.45 pc, which is even smaller
than the extension of the X-rays ($\sim3.5$ pc) assuming a distance
of 10 kpc \citep[][]{Lea09}. A smaller $\eta_{\rm
B}=0.3\times10^{-3}$, and the maximum energy of the particles is set
to 500 TeV to reproduce the observational fluxes in the X-rays and
$\gamma$-rays. The resulting multiband nonthermal emission is
indicated in Fig.\ref{Figj1640b} with soft densities of 1.0 eV
cm$^{-3}$ and 6.0 eV cm$^{-3}$ for the infrared and the optical soft
photons, respectively. In this scenario, the PWN has been compressed
by the reverse shock, and the resulting flux with energies below 1
eV is about two orders of magnitude higher than that in the CASE 1.

\section{Discussion and conclusions}
\label{sec:discussion}

Motivated by the finding that the spectrum of the particles
downstream of a relativistic shock consists of two components: a
relativistic Maxwellian and a power-law high-energy tail with an
index of $-2.4\pm0.1$ \citep[][]{Sp08}, we investigate the
possibility of particles with this new spectrum injected in PWNe
from the TS based on the studies of multiband emission from PWNe.
Following the dynamical method proposed in \citet[][]{GSZ09}, we
study the dynamical and multi-band radiative properties of the three
composite SNRs G0.9+0.1, MSH 15-52 and G338.3-0.0. With appropriate
parameters, we find that a typical PWN is an important $\gamma$-ray
emitter during its evolution although the non-thermal radiation from
the radio to the X-ray band is insignificant sometimes. The
multiband observations of the three PWNe in the remnants can be well
reproduced with the new spectrum of the injected particles.
Therefore, our studies on the dynamical and multiwavelength
radiative properties of PWNe provide evidence of high-energy
electrons/positrons can be injected into a PWN with a Maxwellian
plus a power-law high-energy tail from the TS of the PWN.

In modeling the multiband nonthermal emission from a PWN detected in
the radio, X-ray and $\gamma$-ray bands, particles injected with a
spectrum of a broken power-law are widely used to reproduce the
observed multiwavelength emission \citep[e.g.,][]{Vd06,S08,ZCF08}.
Of course, for the three PWNe discussed in this paper, the multiband
observed spectra of them can also be explained if the particles are
injected with a broken power-law. However, it is unclear why the
broken power-law spectrum is valid when using it to reproduce the
multiwavlength emission from a PWN. From our calculations, we have
found out that the energy distribution of the electrons/positrons in
the nebula can be approximated as a broken power-law with an index
$\sim 1$ in the lower-energy band and an index of $\sim2.5$ in the
higher-energy part before the PWN undergos significant compression,
which is most likely the physical explanation of the broad usage of
a broken power law in modeling the multi-band non-thermal emission
from PWNe.

In this paper, high-energy electrons/positrons are injected into the
PWN from the TS, and the main energy of the nebula is contained in
these particles. These particles undergo radiative and adiabatic
losses when the nebula evolves in the host SNR. Our study indicates
that, for a typical PWN with the parameters similar as G0.9+0.1, the
adiabatic loss of the particles in the nebula is significant after
an age of $\sim1000$ yr (see Fig.\ref{Epwn} and Fig.\ref{Epower}).
Multiwaveband nonthermal emission from a PWN has been investigated
using a simplified time-dependent injection model, in which
high-energy electrons/positrons are injected into the PWN
\citep[e.g.,][]{Vd06,S08,ZCF08}. The pulsar inside the PWN transfers
a part of its spin-down power to the particles with a spectrum of a
broken power-law. In the simplified time-dependent injection model
in \citet[][]{ZCF08}, synchrotron loss of the particles is taken
into account, whereas the adiabatic one is ignored. As a result,
either a relatively smaller initial spin-down power of the pulsar or
a smaller efficiency of the power to the kinetic energy of the
accelerated electrons/positrons is employed in the model. Moreover,
note that in \citet[][]{ZCF08}, an initial spin-down power of
$1\times10^{38}$ erg s$^{-1}$ for MSH 15-52 was used to investigate
the multiband emission from the PWN, which is a factor of 15 smaller
than that used in this paper. Besides the above reasons, the another
main one is a relatively big spin-down time scale of $\sim 5000$ yr,
which is $\propto \dot{E_0}^{-1}$ in the paper, used by
\citet[][]{ZCF08}, whereas in this paper it is adopted to be $500$
yr. The energy released by the pulsar is mainly determined by $E_0
\min \{T_{\rm age}, \tau_0 \}$, and the value in this paper is not
much bigger than that in \citet[][]{ZCF08}. Therefore, the multiband
observed spectra for MSH 15-52 can be reproduced within the two
scenarios even the initial spin-down power of the pulsar is
significantly different.

\begin{acknowledgements}
We are very grateful to the referee for the helpful comments. This
work is partially supported by the Scientific Research Foundation of
Graduate School of Yunnan University, the National Natural Science
Foundation of China (NSFC 10778702, 10803005), a 973 Program
(2009CB824800), and Yunnan Province under a grant 2009 OC.
\end{acknowledgements}


\begin{thebibliography}{99}

\bibitem [Aharonian et al.(2005a)]{A05a}
Aharonian, F. et al. (HESS Collaboration) 2005a, A\&A, 432, L25  %G0.9+0.1

\bibitem [Aharonian et al.(2005b)]{A05b}
Aharonian, F. et al. (HESS Collaboration) 2005b, A\&A, 435, L17   %MSH 15-52

\bibitem[Aharonian et al.(2006)]{A06}
Aharonian, F., et al. (HESS Collaboration) 2006, ApJ, 636, 777

\bibitem[Atoyan \& Aharonian(1996)]{AA96}
Atoyan, A. M., \& Aharonian, F. A. 1996, MNRAS, 278, 525

\bibitem[Blondin, Chevalier \& Frierson(2001)]{BCF01}
Blondin, J. M., Chevalier, R. A., \& Frierson, D. M. 2001, ApJ, 563,
806

\bibitem[Camilo et al.(2009)]{Cet09}
Camilo, F., Ransom, S. M., Gaensler, B. M., \& Lorimer, D. R. ApJ,
700, L34

\bibitem[Caswell et al.(1981)]{Cea81}
Caswell, J. L., Milne, D. K., \& Wellington, K. J. 1981, MNRAS, 195,
89

\bibitem[Dubner et al.(2002)]{Dea02}
Dubner, G. M., Gaensler, B. M., Giacani, E. B., Goss, W. M., \&
Green, A. J. 2002, AJ, 123, 337

\bibitem[Dubner et al.(2008)]{DGD08}
Dubner, G., Giacani, E., \& Decourchelle, A. 2008, A\&A, 487, 1033

\bibitem[Funk et al.(2007)]{Fea07}
Funk, S., et al. 2007, ApJ, 267, 517

\bibitem[Gaensler et al.(2001)]{GPG01}
Gaensler, B. M., Pivovaroff, M. J., \& Garmire, G. P. 2001, ApJ,
556, L107

\bibitem [Gaensler et al.(2002)]{G02}
Gaensler, B. M., Arons, J., Kaspi, V. M., Pivovaroff, M. J., Kawai,
N., \& Tamura, K. 2002, ApJ, 569, 878

\bibitem[Gaensler et al.(1999)]{G99}
Gaensler, B. M., Brazier, K. T. S., Manchester, R. N., Johnston,
S.,\& Green, A. J. 1999, MNRAS, 305, 724

\bibitem[Gaensler \& Slane(2006)]{GS06}
Gaensler, B. M., \& Slane, P. O. 2006, ARA\&A, 44, 17

\bibitem[Gelfand et al.(2007)]{Get07}
Gelfand, J. D., Gaensler, B. M., Slane, P. O., Patnaude, D. J.,
Hughes, J. P., \& Camilo, F. 2007, ApJ, 663, 468

\bibitem[Gelfand et al.(2009)]{GSZ09}
Gelfand, J. D., Slane, P. O., \& Zhang, W. 2009, ApJ, 703, 2051

\bibitem[Goldreich \& Julian(1969)]{GJ69}
Goldreich, P., \& Julian, W. H. 1969, ApJ, 157, 869

\bibitem[Helfand \& Becker(1987)]{HB87}
Helfand, D. J., \& Becker, R. H. 1987, ApJ, 314, 203


\bibitem[Kaplan et al.(2008)]{Ket08}
Kaplan, D. L., Chatterjee, S., Gaensler, B. M., \& Anderson, J.
2008, ApJ, 677, 1201

\bibitem[La Rosa et al.(2000)]{Let00}
La Rosa, T. N., Kassim, N. E., Lazio, T. J. W., \& Hyman, S. D.
2000, AJ, 119, 207

\bibitem[Landi et al.(2006)]{Lea06}
Landi, R., et al. 2006, ApJ,  651, 190

\bibitem[Lemiere et al.(2009)]{Lea09}
Lemiere, A., Slane, P., Gaensler, B. M., \& Murray, S. 2009, \apj,
706, 1269

\bibitem[Livingstone et al.(2005)]{Lea05}
Livingstone, M. A., Kaspi, V. M., Gavril, F. P., \& Manchester, R.
N. 2005, ApJ, 619, 1046

\bibitem[Mereghetti et al.(1998)]{MSI98}
Mereghetti, S., Sidoli, L., \& Israel, G. L. 1998, A\&A, 331, L77

\bibitem[Mineo et al.(2001)]{M01}
Mineo, T. et al. 2001, A\&A, 380, 695

\bibitem[Nakamori et al.(2008)]{Nea08}
Nakamori, T. et al. 2008, ApJ, 677, 297

\bibitem[Proquet et al.(2003)]{PDW03}
Porquet, D., Decourchelle, A., \& Warwick, R. S. 2003, A\&A, 401,
197

\bibitem[Porter et al.(2006)]{Pea06}
Porter, T. A., Moskalenko, I. V., \& Strong, A. W., 2006, ApJ 648 ,
L29

\bibitem[Reynolds \& Chevalier(1984)]{RC84}
Reynolds, S. P. \& Chevalier, R. A., 1984, ApJ, 278, 630

\bibitem [Seward et al.(1982)]{S82}
Seward, F. D., \& Harnden, F. R., Jr. 1982, ApJ, 256, L45

\bibitem[Seward et al.(1983)]{Sea83}
Seward, F. D., Harnden, F. R., Jr., Murdin, P., \& Clark, D. H.
1983, ApJ, 267, 698

\bibitem[Sidoli et al.(2000)]{Set00}
Sidoli, L., Mereghetti, S., Israel, G. L., \& Bocchino, F. 2000,
A\&A, 361, 719

\bibitem[Slane et al.(2008)]{Set08}
Slane, P., Helfand, D. J., Reynolds, S. P., Gaensler, B.M., Lemiere,
A., \& Wang, Z. 2008, ApJ, 676, L33

\bibitem[Slane(2008)]{S08}
Slane, P. 2008, AIPC, 1085, 120

\bibitem[Spitkovsky(2008)]{Sp08}
Spitkovsky, A. 2008, ApJ, 682, L5

\bibitem[Strong et al.(2000)]{SMR00}
Strong, A. W., Moskalenko, I. V., \& Reimer, O. 2000, ApJ, 537, 763

\bibitem[Sugizaki et al.(2001)]{Sea01}
Sugizaki, M., et al. 2001, ApJS, 134, 77

\bibitem[Taylor \& Cordes(1993)]{TC93}
Taylor, J., H. \& Cordes, J. M. 1993, ApJ, 411, 674

\bibitem[Trussoni et al.(1996)]{Tea96}
Trussoni, E., Massaglia, S., Caucino, S., Brinkmann, W., \&
Aschenbach, B. 1996, A\&A, 306, 581

\bibitem[Truelove \& McKee(1999)]{TM99}
Truelove, J. K., \& McKee, C. F. 1999, ApJS, 120, 299

\bibitem[van der Swaluw et al.(2001)]{vet01}
van der Swaluw, E., Achterberg, A., Gallant, Y. A., \& T\'{o}th, G.
2001, A\&A, 380, 309

\bibitem[van der Swaluw et al.(2004)]{vDK04}
van der Swaluw, E., Downes, T. P., \& Keegan, R. 2004,

\bibitem[Venter \& de Jager(2006)]{Vd06}
Venter, C., \& de Jager, O. C. 2006, in Proc. 363rdWE-Heraeus
Seminar, Neutron Stars and Pulsars, ed. W. Becker \& H. H. Huang
(MPE Report 291; Garching: MPE), 40

\bibitem[Volpi et al.(2008)]{Vet08}
Volpi, D., Del Zanna, L., Amato, E. \& Bucciantini, N. 2008, A\&A,
485, 337

\bibitem[Whiteoak \& Green(1996)]{WG96}
Whiteoak, J. B. Z., \& Green, A. J. 1996, A\&AS, 118, 329

\bibitem[Zhang et al.(2008)]{ZCF08}
Zhang, L., Chen, S. B., \& Fang, J. 2008, ApJ, 676, 1216



\end{thebibliography}
\end{document}